%
\pretolerance=10000  

%
%
%

\def\o{{\rm o}}

\def\B{{\bf B}}
\def\b{{\bf b}}
\def\Bo{B_{\rm o}}
\def\Bt{{B_{\rm t}}}
\def\Bu{{B_{\rm u}}}

\def\e{{\bf e}}
\def\etal{et al.}
\def\Gammad{\Gamma_{\rm d}}
\def\Gammas{\Gamma_{\rm s}}
\def\gameff{\gamma_{\rm eff}}
\def\kb{{\bf k}_B}
\def\nic{n_{\rm IC}}
\def\Peq{P_{\rm eq}}
\def\rhoc{\rho_{\rm c}}
\def\rhoic{\rho_{\rm IC}}

\def\Teq{T_{\rm eq}}
\def\u{{\bf u}}
\def\va{v_{\rm A}}
\def\vBo{{\bf B}_{\rm o}}

\def\x{{\bf x}}

\def\VS{V\'azquez-Semadeni}

%
\def\ltsima{$\; \buildrel < \over \sim \;$}    
\def\simlt{\lower.5ex\hbox{\ltsima}}           
\def\gtsima{$\; \buildrel > \over \sim \;$}    
\def\simgt{\lower.5ex\hbox{\gtsima}}           

\def\doublespace {\smallskipamount=6pt plus0pt minus0pt
                  \medskipamount=12pt plus0pt minus0pt
                  \bigskipamount=24pt plus0pt minus0pt
                  \normalbaselineskip=24pt plus0pt minus0pt
                  \normallineskip=2pt
                  \normallineskiplimit=0pt
                  \jot=6pt
                  {\def\smallskip {\vskip\smallskipamount}}
                  {\def\medskip   {\vskip\medskipamount}}
                  {\def\bigskip   {\vskip\bigskipamount}}
                  {\setbox\strutbox=\hbox{\vrule
                    height17.0pt depth7.0pt width 0pt}}
                  \parskip 12.0pt
                  \normalbaselines}

\def\singlespace {\smallskipamount=3pt plus1pt minus1pt
                  \medskipamount=6pt plus2pt minus2pt
                  \bigskipamount=12pt plus4pt minus4pt
                  \normalbaselineskip=12pt plus0pt minus0pt
                  \normallineskip=1pt
                  \normallineskiplimit=0pt
                  \jot=3pt
                  {\def\smallskip {\vskip\smallskipamount}}
                  {\def\medskip   {\vskip\medskipamount}}
                  {\def\bigskip   {\vskip\bigskipamount}}
                  {\setbox\strutbox=\hbox{\vrule
                    height8.5pt depth3.5pt width 0pt}}
                  \parskip 0pt
                  \normalbaselines}

\magnification=\magstephalf
\hsize 5.5truein
\vsize 8.0truein
\hoffset .5truein
\voffset 0.5truein
%
\parindent 24.0pt

\doublespace

\baselineskip=18truept
\singlespace


\centerline{\bf A TURBULENT MODEL FOR THE INTERSTELLAR MEDIUM.}
\centerline{\bf II. MAGNETIC FIELDS AND ROTATION}
\smallskip
\centerline{Thierry Passot$^1$, Enrique V\'azquez-Semadeni$^2$, and
Annick  Pouquet$^1$}
\bigskip
\centerline{$^1$Observatoire de Nice, BP 229, 06304 Nice CEDEX 4,
FRANCE}
\centerline{I: passot@obs-nice.fr, pouquet@obs-nice.fr}
\smallskip
\centerline{$^2$Instituto de Astronom\'{\i}a, UNAM, Apdo. Postal 70-264}
\centerline{M\'exico D. F., 04510, MEXICO.}
\centerline{I: enro@astroscu.unam.mx}
\bigskip
\centerline{\bf Abstract}
\noindent

We present results from two-dimensional numerical simulations of a supersonic
turbulent flow with parameters characteristic of the interstellar
medium at the 1 kpc scale in the plane of the galactic disk, incorporating
shear,
thresholded and discrete star formation (SF), self-gravity, rotation and
magnetic
fields.
A test of the model in the linear regime supports the results of the linear
theory of Elmegreen (1991a). At low shear, a weak azimuthal
magnetic field stabilizes
the medium by opposing collapse of radial perturbations, while a strong field
is destabilizing by preventing Coriolis spin-up of azimuthal perturbations
(magnetic braking). At high shear, azimuthal perturbations are sheared into the
radial direction before they have time to collapse, and the magnetic field
becomes stabilizing again.

In the fully nonlinear turbulent regime, while some results
of the linear theory persist, new effects also emerge.
The production of turbulent density fluctuations appears to be affected by the
magnetic field as in the linear regime: moderate field strengths cause a
decrease in the time-integrated star formation rate,
while larger values cause an increase. A result not predicted by the linear
theory
is that for very large field strengths,
a decrease in the integrated SFR obtains again, indicating
a ``rigidization'' of the medium due to the magnetic field.
Other exclusively nonlinear effects are:
a) Even though there
is no dynamo in 2D, the simulations are able to
maintain or increase their net magnetic energy in the presence of a seed
uniform azimuthal component.
b) A well-defined power-law magnetic spectrum and an inverse magnetic cascade
are observed in the simulations, indicating full MHD turbulence.
Thus, magnetic field energy is generated
in regions of SF and cascades up to the largest scales.
c) The field has a slight but noticeable tendency to be
aligned with density features. This appears to be as much a consequence of
the gas pushing on the magnetic field as due to constraints on gas motions
because
of the presence of the magnetic field.
d) A ``pressure cooker'' effect is observed in
which the magnetic field prevents HII regions from expanding freely, as in the
recent results of Slavin \& Cox (1993).
e) The orientation of the large-scale azimuthal field appears to follow that of
the large-scale Galactic shear.
f) A tendency to exhibit {\it less} filamentary structures at stronger values
of the uniform component of the magnetic field
is present in several magnetic runs. Possible mechanisms that may lead to
this result are discussed.
g) For fiducial values of the parameters, the flow in general appears to be in
rough equipartition between magnetic and kinetic energy.
There is  no clear domination of either the magnetic or the inertial forces.
h) A median value of the magnetic field strength within clouds is $\sim
12\mu$G,
while for the intercloud medium
a value of $\sim 3\mu$G is found. Maximum contrasts of up to a
factor of $\sim 10$ are observed.

\medskip
\noindent
{\it Subject headings:} ISM: clouds -- ISM: evolution -- ISM: magnetic fields
-- ISM: structure -- instabilities -- turbulence.

\medskip
\noindent
Appeared in {\it The Astrophysical Journal}, 455, 536.
\vfill
\eject

\centerline{1. INTRODUCTION}

In Paper I of this series (V\'azquez-Semadeni, Passot \& Pouquet 1995; see
also Passot, \VS\ \& Pouquet 1994),
we presented two-dimensional (2D) numerical simulations attempting to model the
Interstellar Medium (ISM) in the plane
of the Galaxy at the kpc scale. The model incorporates
self-gravitating hydrodynamics,
parameterized radiative cooling, diffuse heating, and a prescription for
modeling star formation (SF). Some of the first results obtained from the
numerical simulations were
the ``slaving'' of the temperature and thermal pressure to the density in the
absence of thermal instabilities due to the short thermal time scales, the
existence of a self-sustained cycle in which the stellar energy input to the
ISM is enough to maintain the turbulence (with an efficiency of 0.05 \%), the
formation of clouds and cloud complexes mainly through collisions of gas
streams (turbulent ram pressure), and a large scatter in the virial ratios of
cloud energies, although nearly-virial clouds exhibited a tendency to live
longer,
explaining their observed overabundance.

Among the significant shortcomings of the model of Paper I,
the most notorious one is an excessive
cloud temperature ($\simgt 1000$ K), and consequently, too low a density
contrast ($\rho_{\rm max}/\rho_{\rm min}\sim 50$). Additionally, two of the
most
obvious omissions are magnetic fields and rotation of the Galactic disk.
In the present paper we present improvements of the model along both lines:
we first introduce a different diffuse heating mimicking the effect of
shielding
against background UV radiation; this allows the flow to
reach more realistic cloud temperatures (a few hundred K) and density contrasts
($\sim 1000$). We then extend the model by incorporating both magnetic fields
and rotation.

The magnetic field in the ISM is ubiquitous,
although its precise dynamical effects (e.g. Troland 1990) and topology
(e.g. Trimble 1990; Heiles et al.\ 1993) are still a matter of active research
and debate. Some fundamental unanswered questions concerning the field are:

\parindent=0pt
1) Is the field of primordial origin or is it continually generated and
dissipated in the Galaxy? In the latter case, what
are the generation/amplification mechanisms? (e.g.\ Zweibel 1987;
Wielebinsky \& Krause 1993;
Tajima, Cable, Shibata \& Kulsrud 1992)?

2) What is the amplitude ratio between its uniform and
fluctuating components, and what are the typical scales of the latter?

3) Is the field fully turbulent, or does it consist
simply of a superposition of weakly interacting waves?

4) What is its relative importance in the global dynamics and energetics
of the ISM and in
cloud formation and support (e.g., Falgarone \& Puget 1986, Troland 1990,
Pudritz \& G\'omez de Castro 1991; Elmegreen 1991a; McKee et al.\ 1993)?

5) Does the field strength correlate with gas density (Mouschovias, 1976a, b;
Garc\'\i a-Barreto
et al.\ 1987; Crutcher, Kazes \& Troland 1987; Myers \& Goodman 1988a)?

6) Does the field orientation correlate with density features (cloud
shapes and elongations) (Goodman 1991; see also Heiles et al.\ 1993)?

7) Is there a tendency for equipartition between kinetic, magnetic and
gravitational energies in molecular clouds (MCs) as observations seem
to indicate (Myers \& Goodman 1988a,b)?

8) Is the Alfv\'en speed the typical velocity of propagation of disturbances
in the ISM (Falgarone \& Puget 1986; Myers \& Goodman 1988a,b)?

\parindent=20pt
Concerning rotation, it has long been known that it stabilizes the
gravitational collapse of structures in the Galactic disk
(Chandrasekhar, 1961;
Toomre 1964; Goldreich \& Lynden-Bell 1965). Two main
effects are present: the conversion of compressive motions into shearing ones
by the Coriolis force, and the ``restoring force'' against radial motions
arising from the interplay between the radial component of gravity and the
centrifugal force.  Recently,
a combined instability analysis of a self-gravitating, rotating flow with
heating and cooling and magnetic fields has been performed by Elmegreen (1991a,
1994). In the turbulent case, the main mode of cloud formation identified in
Paper I---that arising from random turbulent compressions---may
also be inhibited by rotation.

In the present paper, we present two-dimensional numerical simulations aimed at
investigating these problems. Our simulations
are limited by two facts. First, since the simulations reported are 2D,
they cannot produce a dynamo, which is intrinsically a 3D
effect. However, this has the advantage that other mechanisms of amplification
of the magnetic field can be isolated and identified. Second, the numerical
code used is pseudospectral, with periodic boundary conditions. The latter
imply that the centrifugal and radial gravitational forces, which depend on the
galactocentric distance, cannot
be included in the model. However, as
shown in \S 2.2, all the relevant aspects of the dynamics are
included in the Coriolis force and the large scale shear,
so that the omission turns out to be inconsequential.

The plan of the paper is as follows: \S 2 presents the equations of the model
and discusses the values of the parameters; \S 3 describes the
results of simulations in the linear regime, and \S 4 discusses the nonlinear
behavior of the model, stressing in particular
the differences between the two regimes.
In \S 5 we discuss the general behavior of a simulation with fiducial values of
the parameters. Finally, \S 6
contains a brief summary and a discussion of the reaches and limitations of the
current simulations.

\bigskip
\centerline{2. THE MODEL}

\medskip
\centerline{2.1. {\it Equations}}

As in Paper I, we use a single-fluid approach to represent the ISM with
several source terms in order to model radiative cooling, large-scale shear,
and
stellar and diffuse heating. The computations solve the
nondimensionalized equations
$${\partial\rho\over\partial t} + {\nabla}\cdot (\rho\u) = \mu
\nabla^2 \rho, \eqno(1a)$$
$${\partial\u\over\partial t} + \u\cdot\nabla\u =
-{\nabla P\over \rho} - {\nu_8} {\nabla^8\u}
- \Bigl({J \over M_a}\Bigr)^2 \nabla \phi + {1 \over \rho}
\bigl(\nabla \times {\bf B}\bigr) \times {\bf B} - 2 \Omega \times \u,
\eqno(1b)$$
$${\partial e\over\partial t} + \u\cdot\nabla e = -(\gamma -1)
e\nabla\cdot \u + {\kappa}_T {\nabla^2e\over \rho} + \Gammad
+ \Gammas - \rho \Lambda, \eqno(1c)$$
$${\partial {\bf B}\over\partial t} = \nabla \times (\u \times {\bf B})
- {\nu_8} {\nabla^8{\bf B}}, \eqno(1d)$$
$$\nabla^2 \phi=\rho -1, \eqno(1e)$$
in two dimensions [on the $(x,y)$-plane with $\partial/\partial z = 0$]
using periodic boundary conditions. As
usual, $\rho$ is the density, {\bf u} is the fluid velocity, $e$ is the
internal energy per unit mass, $P$ is the thermal pressure,
\B\ is the magnetic field,
$\Omega$ is the angular velocity of the rotation,
and $\phi$ is the gravitational potential.
Furthermore, we will frequently make use of the number density $n \equiv \rho /
m_{\rm H}$, where $m_{\rm H}$ is the mass of the hydrogen atom. We
recover identically the model of Paper I by setting \B\
 $ \equiv 0$
and $\Omega \equiv 0$
in the above equations, although some of the parameter values may have changed,
and the diffuse heating is different (see below, \S 2.2).

We use an ideal-gas equation of state $P=(\gamma-1)\rho e$,
where $\gamma = c_p/c_v$ is the ratio
of specific heats at constant pressure and volume, respectively. We take
$\gamma=5/3$. The temperature is related
to the internal energy by $e =c_vT$.  The variables are normalized
to characteristic values $\rho_\o, u_\o, L_\o$, $T_\o$
and $\Bo$,
given in Table 1. We refer the reader to Paper I for a
thorough discussion of most of the parameters and model terms.
Section 2.2 below describes those that are
new or have changed for the present paper.
The physical dimension corresponding to the side
of the integration box is $L_\o$. The nondimensional parameters resulting from
the normalization are: the Mach number $M_a= u_\o/c_\o$, where $c_\o =
\bigl(\gamma
kT_\o/m_H\bigr)^{1/2}$ is the adiabatic speed of sound at the normalizing
temperature $T_\o$; the Jeans number $J = L_\o/L_J$, giving
the number of Jeans' lengths $L_J = (\pi c_\o^2/G \rho_\o)$ in the simulation.
The thermal diffusivity is denoted $\kappa_T$.

The pseudospectral scheme used here introduces no numerical
viscosity, since in Fourier space the MHD equations become a set of coupled
ordinary differential equations for the Fourier amplitudes.
Thus, we include explicit dissipation terms in the equations for \u\ and \B.
A hyperviscosity scheme with
a $\nabla^8$ operator is used, which confines viscous
effects to the smallest resolved scales.
Indeed, for a fixed amount of dissipation at a given small scale, the higher
power
of the Laplacian allows for a smaller  effective dissipation at the large
scales.
This technique is widespread in the fluid dynamics community (see, for
example, McWilliams 1984; Babiano \etal, 1987). Note that the diffusion
coefficients for the velocity and magnetic fields are taken to be the
same and equal to $\nu_8$.

In addition,
a mass diffusion $\mu\nabla^2\rho$ is included in the continuity equation
in order to smooth out the density gradients, thus allowing the simulations to
reach higher rms Mach numbers. The effects of this term are discussed at length
in Paper I.

Note that the kinematic viscosity and the
mass diffusion coefficients are chosen so that velocity and density
discontinuities are spread out a few pixels, guaranteeing that they can be
resolved.
Also, note that heat generated by the kinetic and
magnetic dissipations
is not included in the calculations because it is negligible compared to the
stellar and diffuse heating.

Concerning self-gravity, we note that
equation (1e) is a modified Poisson equation appropriate for infinite media,
and its application to pseudospectral simulations with periodic boundary
conditions has been discussed in detail
by Alecian \& L\'eorat (1988). Essentially, this equation
represents the self-gravity of the density {\it fluctuations}.

Further details on the pseudospectral integration technique we use can be
found in Paper I. The theoretical bases on which it rests can be found in
Canuto et al.\ (1988).
As is standard for turbulence simulations, the initial conditions for all
variables are Gaussian fluctuations with random phases.
The typical scale
size of the initial density, temperature and velocity fluctuations
is 1/8 of the integration
box. The typical scale size of the initial magnetic field fluctuations is
specified by the parameter $\kb$, discussed in \S 2.3.2 below.

A few words are in order concerning the
choice of units (see Table 1). The unit of length $L_\o$, equal to 1 kpc in
physical units, is $2 \pi$ in code units. Derived units which
involve this length also contain factors of $2 \pi$. Also,
the unit of velocity
$u_\o$ is chosen equal to the speed of sound $c_\o$ at the unit of temperature
$T_\o$. The fundamental density unit in the code has units of a
volume density.
Finally, we note that the  time step in a typical run
varies in the range .01--.001 in units of $1.3 \times 10^7$ years.

\centerline{2.2. {\it Model Terms}}

In this section we first give a brief summary of the forms used for the model
terms. Only model terms
that are new or have changed from Paper I are discussed in detail here.
A detailed description of all other terms can be found in Paper I.
Numerical values of their associated parameters are given in Table 1.

\medskip
\centerline{2.2.1. {\it Diffuse heating $\Gammad$}}

As pointed out
in the Introduction, the simulations
presented in Paper I had one important shortcoming: the densest regions
that can be associated with clouds still had temperatures much larger
($\sim 1,500$ K) than actual
cloud temperatures (10--100 K) together with too low a density ($\rho_{\rm max}
\sim 10 \rho_\o$).
This problem was mainly due to the form of the
diffuse heating used in that paper, namely a constant, uniform heat source.
This form of the diffuse heating produced realistic temperatures for
the Intercloud Medium (ICM) (although see the discussion in Paper I concerning
the uncertainties surrounding the temperature of this $n \sim 1$ cm$^{-3}$
gas), but it caused the medium to be too ``hard'' (resistive to compression),
thus
both reducing the density contrast ($\rho_{\rm max}/\rho_{\rm min} \sim 50$),
and
keeping the cloud temperatures too high. Physically, if most of the diffuse
background energy is in low-energy UV photons, then it is known that
clouds can shield their interiors from this radiation if their column densities
are high enough (Franco \& Cox 1986). Conversely, if most of the background
energy is in energetic particles, then the heating rate per unit mass
should be rather insensitive to the cloud density. As a compromise between
these
two possibilities, we now adopt a new diffuse heating function of the form
$$\Gammad({\bf x},t)={\Gamma_\o(\rho/\rhoic)^{-\alpha}}$$
where $\rhoic$ is a typical density of the ICM and $\alpha$ is a free
parameter,
which in general we take such as to give a weak dependence of $\Gammad$ on
$\rho$.
The factor $\rhoic^{-1}$ in the density is introduced so that an equilibrium
temperature (between diffuse heating and radiative cooling) of $10^4$ K is
obtained at the density of the ICM.
This form of $\Gammad$ is not intended to represent any
realistic physical dependence, since
self-shielding depends on column density, which is a non-local cloud property,
and thus extremely costly to implement in our code. Instead, the adopted form
provides a smooth density dependence. An experiment with a threshold
criterion for turning off the diffuse heating above a critical density was
presented in Paper I, but it was found that it introduced spurious oscillations
of the density, temperature and pressure distributions in clouds.
The fiducial values we adopt for the parameters of the diffuse heating
are $\nic \equiv \rhoic/m_{\rm H} = 0.2$
cm$^{-3}$ and $\alpha=1/2$. A discussion of the effects of variations
in $\alpha$ (which is reflected on an effective polytropic exponent $\gameff$
[\S 3.1]) on the stability of the model is given in \S 3.2.

\medskip
\centerline{2.2.2. {\it Stellar Heating $\Gammas$} }

We model local stellar heating due to massive
stars by means of a threshold algorithm: a heating center a few pixels across
(with a Gaussian profile)
is turned on wherever the density exceeds a threshold value $\rhoc$ and $\nabla
\cdot \u < 0$. Once a ``star'' is turned on, it stays on for a time $\Delta t =
6 \times 10^6$ yr, typical of the lifetime of OB stars. Note that once a star
is turned on, it remains fixed with respect to the numerical grid.

As described in Paper I, the star formation rate (SFR) is
computed in the simulations as the fraction of pixels that reach
$\rhoc$ per unit time.

\medskip
\centerline{2.2.3. {\it Cooling $\Lambda$}}

We use the parameterization of the radiative
cooling functions of Raymond \& Cox (1976)
and Dalgarno \& McCray (1972), as employed by Rosen \etal\ (1993) and Rosen \&
Bregman (1994):
$$\Lambda = \cases {0          & $ \ \ \ 0 \le T < 100~K$ \cr
                       \Lambda_1 T^2       & $ \ 100~K \le T < 2000~K$ \cr
                       \Lambda_2 T^{1.5}   & $2000~K \le T < 8000~K$ \cr
                       \Lambda_3 T^{2.867} & $8000~K \le T < 10^5$K\cr
                       \Lambda_4 T^{-0.65} & $ \ \ 10^5~K \le T < 4 \times
10^7~K$\cr} $$

Similarly to Paper I, the cooling and heating rates are decreased by a factor
of 7
with respect to realistic values for numerical reasons, in order to reduce the
stiffness of the resulting system. This, however,
is not expected to affect the dynamics as their characteristic time scales are
still much shorter than dynamical time scales (by factors between
10 and $10^4$).
As discussed in Paper I, the shorter thermal time scales have the interesting
consequence that the system is capable to reach thermal equilibrium between
cooling and diffuse heating in times much shorter than the characteristic time
for turbulence-induced density fluctuations
to occur. The flow thus behaves as a polytropic gas
with an effective polytropic exponent given by the condition of thermal
equilibrium (\S 3.1). This behavior is preserved as long as the thermal
rates are much faster than the dynamical rates, so that the equilibrium
temperature can be reached before the density changes appreciably, as is the
case of our simulations.
Even in the case of strong heating due to the localized
stellar heating, thermal equilibrium between the latter and the cooling is
achieved virtually instantaneously. In this case a strong pressure
imbalance with the surroundings exists, due to the pointwise nature of the
stellar heating, producing expanding motions.
Thus, the dynamical effects of the stellar heating are
preserved in spite of the reduced thermal rates we use.

It should also be pointed out that in our simulations the gas temperature never
exceeds $3.5 \times 10^4$ K, since heating due to massive-star winds
and supernovae is not considered (see \S 6.2 for a discussion of limitations of
the simulations). Thus, the strong thermal instability above $10^5$ K in the
cooling function is never reached in practice, and is included only for
completeness. Work including such heating mechanisms is in progress (Gazol et
al. 1995).

Finally, the lower cutoff of the cooling is discussed at length in Paper I,
although note that here we have extended it to 100 K, down from 300 K in Paper
I. This limit still does not pose significant numerical difficulties, yet it
allows the simulations to achieve more realistic cloud temperatures.

\medskip
\centerline{2.2.4. {\it Rotation}}

We assume that our integration domain is located roughly
at the solar circle, rotating about the Galactic center at a radius $R_\o$ and
with angular speed $\Omega_\o$, with the $x$- and
$y$-coordinates in the model corresponding to the azimuthal and radial
directions, respectively.
In this rotating frame one must consider the Coriolis
force $-2 \Omega_\o \times \u$, the centrifugal force $\Omega_\o^2 R \hat {\bf
e}_y$ and the radial gravitational force ${\bf g}_R$
of the total mass contained within radius $R$.
In the actual Galactic disk, the latter two
forces exactly balance each other at every
radius R, and ${\bf g}_R = - \Omega^2(R) R \hat {\bf e}_y$. This balance is at
the origin of the differential rotation of the Galactic disk and its associated
shear. In a rotating frame characterized by a single angular velocity
$\Omega_\o$, the total fluid velocity is the sum of the shear
$\u_{\rm s} = u_{\rm s} \hat {\bf e}_x$ and a turbulent fluctuation
$\u_{\rm turb}$. Then,
at radius $R$, the difference between the centrifugal force  due to the
rotation velocity of the frame and the radial gravitational
force is, to first order,
$-2 \Omega_\o (d\Omega/dR)_{R_\o} R y \hat {\bf e}_y$, and this excess is
balanced by the Coriolis force acting on the shear velocity, which gives
$u_{\rm s} = - R (d \Omega/dR) y$.

The simulation of this interaction of forces and the resulting shear can be
accomplished at two equivalent levels. At a more fundamental level, the radial
gravitational force can be included in the equations, with an explicit $R$-
(or, equivalently, a $y$-) dependence. At a less physical, but less numerically
costly level, the resulting shearing profile can be simply imposed on the
velocity field, without introducing the radial centrifugal force in the
equation of motion. We have chosen the latter option. Note, however, that
due to the periodic boundary conditions we use, we cannot introduce a
monotonic function of radius. Instead, we need to introduce
a periodic function in the integration box, the simplest of which is a
sinusoidal profile of period one of the form
$$u_x(y)=A_\o \sin {{2\pi y}\over L_\o}.\eqno(2)$$
This is trivially accomplished by fixing the first Fourier mode of the velocity
to a constant. We choose the shear to have period one in order to approximate
in as large as possible a fraction of the box the monotonic character of the
actual shear in the Galactic disk.

It should be kept in mind that,
as given by the above expression, the shear has the
correct sign of $d\Omega/dR$ (i.e., $\Omega$ decreasing outwards)
in only half of the integration box (the upper
and lower quarters of the box), and the opposite sign in the central half
of the box. This has dynamical consequences which will be
discussed in \S 3.2.

Toh, Ohkitani \& Yamada (1991)
have developed an algorithm for simulating incompressible 2D flows with
linear shear profiles with pseudospectral techniques (see also
Feireisen, Reynolds \& Ferziger 1981 for the compressible case). In MHD,
a special choice of gauge for the magnetic potential can be used to accommodate
shear profiles (Brandenburg et al.\ 1995).
Here, for simplicity, we have adopted the above sinusoidal velocity
profile.

\medskip
\centerline{2.3. {\it Parameters}}

A detailed description of the criteria used to select most parameter values
is given in Paper I. Table 1 gives a summary of the parameters' meanings
and fiducial values, both in physical units and in nondimensional code
units. In this section we only
discuss the values and criteria for the new
parameters related to the magnetic field and the Coriolis force.

\smallskip
\centerline{2.3.1 {\it Rotation}}

We adopt a rotational speed of 250 km s$^{-1}$ (see, e.g., Shore
1989), implying $\Omega=
3.14 \times 10^{-8}$ yr$^{-1}$, which, in units of
the code becomes $\Omega= 0.41$.

\smallskip
\centerline{2.3.2. {\it Magnetic field}}

Current views picture the Galactic magnetic field along the Galactic plane
in the solar neighborhood as having a nearly azimuthal uniform component
of strength $\Bu \sim 1.5 \mu$G, and a turbulent component $\Bt \sim 5 \mu$G
(e.g. Rand \& Kulkarni 1989, Wielebinski \& Krause 1993).
We choose a unit of the magnetic field strength such that
$\va = u_\o$.
At the unit of density for the code ($\rho_\o = 1$ cm$^{-3}$), this corresponds
to $\Bo = 5\mu$G.
Note that, in the nondimensional code units the Alfv\'en
speed is given by $v_{\rm A}^2 = B^2/\rho$.

Finally, as mentioned in \S 2.1, another initial parameter of the simulations
is $\kb$, the characteristic wavenumber of the initial magnetic field
fluctuations. We adopt a fiducial value $\kb=1$, although the effect
of varying this parameter is described in \S 4.2.

\bigskip
\centerline{3. LINEAR EVOLUTION}

As will be seen in \S 4, the fully nonlinear behavior of the model is extremely
complex. In order to appropriately interpret the various processes at play and
to distinguish between linear and nonlinear effects, we first discuss the
linear evolution of the model. The discussion in this and the following
sections relies heavily on the results of nearly 70 runs we performed,
which we used to explore
the effects of variations in the parameters (low resolution runs, typically
$128^2$ grid points) and to analyze the structure and evolution
of the model with the fiducial values of
the parameters (run 28, with $512^2$ grid points). A summary of the
parameters of the runs referred to in this paper is given in Table 2.

\medskip
\centerline{3.1. {\it Equilibrium state}}

For the linear evolution runs, the system is started from an equilibrium state
with uniform density $n_\o =1$ cm$^{-3}$, velocity field given by eq.\
(2), uniform temperature and a uniform magnetic field in the $x$
(``azimuthal'') direction. Several cases with various
values of the initial uniform
magnetic field have been analyzed. The initial value of the temperature is that
corresponding to equilibrium between cooling and the diffuse heating, and is
thus a function of the density exponent $\alpha$ in the expression for
$\Gammad$ (\S 2.2). Indeed,
setting $\rho \Lambda = \Gammad$ gives the following equilibrium values for the
temperature and thermal pressure of the flow:
$$\Teq = \Bigl[{\Gamma_\o \rhoic^\alpha \over \Lambda_{\rm i} \rho^{1 +
\alpha}}\Bigr]^{1/\beta_{\rm i}}\ \ \ \ \Peq={\rho \Teq \over \gamma} =
{\rho^{\gameff}\over \gamma}
\Bigl[{\Gamma_\o \rhoic^\alpha \over \Lambda_{\rm i}}\Bigr]
^{1/\beta_{\rm i}},\eqno(3)$$
where $\gameff= 1-{(1+\alpha) / \beta_{\rm i}}$, and $\beta_{\rm i}$ and
$\Lambda_{\rm i}$ are respectively the exponent and coefficient of the
temperature in the $i$-th range of the cooling function. Thus, the initial
temperature listed in Table 2 for each run, if different from $10^4$ K, is the
equilibrium temperature appropriate for the value of $\alpha$ in that run.
Note that runs in the
fully nonlinear regime, discussed in \S 4, always start at $T=10^4$ K.
For small perturbations about this equilibrium state,
the effective speed of sound $c_{\rm eff}$ satisfies,
in nondimensional units, $c_{\rm eff}^2 = \gameff \Peq
/\rho_{\rm eq} = (\gameff/\gamma) \Teq$.\footnote{$^1$}{Note that in Paper I
the quantity $c_{\rm eff}^2$ used in the dispersion
relation for gravito-acoustic waves was mistakenly written without
the factor $1/\gamma$. This caused an underestimation of the computed
periods, which are then in not as good an agreement with the periods
observed in the simulations. However, the observed and computed periods
still agree within a factor $30\% $.}

\medskip
\centerline{3.2. {\it Dispersion relation and stability criteria}}

The original gravitational instability analysis of Jeans (1902) has been
continually extended by a number of authors to include additional processes,
such as rotation (Chandrasekhar 1961; Toomre 1964; Goldreich \& Lynden-Bell
1965), a variety of energy sources and sinks (Struck-Marcell \& Scalo 1984),
and
all of the above plus magnetic fields (Elmegreen 1991a, 1994; we will refer to
the latter as E94).
In particular, the
system considered by Elmegreen is nearly identical to the system we consider in
our simulations, except that our model is two-dimensional (note also the
reversed choice of the $x$- and $y$-axes).
For the combined instability, E94 gives dispersion relations at $t=0$ for
radial and azimuthal perturbations (see also Elmegreen 1991b).
They read, respectively:
$$\omega_R^2 = 2 \pi G \sigma k - k^2 ({\gameff\over\gamma} c^2 + \va^2) -
\kappa^2 ,\eqno(4a)$$
$$\omega_A^2 = 2 \pi G \sigma k - k^2 {\gameff\over\gamma} c^2 - {\omega_A^2
\kappa^2 \over
\omega_A^2 + k^2 \va^2},\eqno(4b)$$
where $\omega_R$ ( resp.\ $\omega_A$) is the growth rate in
the radial (resp.\
azimuthal) case, $k$ is the wavenumber, $\sigma$ is the surface
density of the disk, $\kappa = 2 \Omega [1 + 1/2\ (R/\Omega)\ d
\Omega/ dR]^{1/2}$ is the epicyclic frequency, and $\va$ is the
Alfv\'en velocity.
Note that in E94 there is no factor of $1/\gamma$ in the second term of
the r.h.s., because in that paper $c$ is defined as the isothermal
sound speed, whereas in the present paper
it is defined as the adiabatic sound speed.
Also, in transcribing eqs.\ (4) from E94,
we have omitted a reduction factor included there to account for
the finite thickness of the disk.
Note also that in E94 $\gameff$ is {\it stricto
sensu} a different quantity than that used in the present paper.
While ours is an {\it equilibrium} $\gameff$ (denoted
$\gameff^{\rm eq}$), in E94 it is a {\it perturbation} $\gameff$
(denoted $\gameff^{\rm p}$), which is obtained from a linear
stability analysis of the internal energy equation, treated separately
(see also Elmegreen 1991b). However, it can be easily shown that in the
limit $\omega \rightarrow 0$,
which corresponds to the cooling rate being much faster
than the growth rate of the perturbation,
$\gameff^{\rm p} \rightarrow \gameff^{\rm eq}$.
This is precisely the
case in our simulations (cf. Paper I). Thus, it is justified to
use $\gameff$ as defined in \S 3.1 in the dispersion relation (4).
Incidentally, in the opposite limit, $ \omega \rightarrow \infty$,
$\gameff^{\rm p} \rightarrow \gamma$, where $\gamma$ is the actual
heat capacity ratio of the gas. In this case, the
growth rate of the perturbation is much faster than the cooling
rate, and the flow responds adiabatically.

For our 2D system without shear ($\kappa = 2 \Omega$), the dispersion relations
for the radial and azimuthal growth rates read, in nondimensionalized form,
$$\lambda_{\rm R}^2 = J^2 - k^2 ({\gameff\over\gamma}T_{\rm eq} + \Bo^2) -
\kappa^2 ,\eqno(5a)$$
$$\lambda_{\rm A}^2 = J^2 - k^2 {\gameff\over\gamma} T_{\rm eq} -
{\kappa^2 \lambda_{\rm A}^2\over \lambda_{\rm A}^2 + k^2 \Bo^2} \eqno(5b)$$
Note that, in contrast to eqs.\ (4), the gravitational term does not
contain the factor $k$, because of the strict two-dimensionality of our model.
When shear is included ($\kappa$ given by the full expression above), these
growth rates cannot be defined any more except for $t \sim 0$.

We now analyze equations (5) in the absence of shear (hence
$\kappa \equiv 2\Omega$). In our 2D system, radial perturbations are
always stable at all scales when $\tilde Q_{\rm R}=\kappa /J >1$
(in the nondimensionalized units of the paper, the volume density
$\langle \rho \rangle$ is equal to unity), whereas in E94 the criterion reads
$$\tilde Q_{\rm R}={{\kappa (c^2{\gameff\over\gamma}+v_{\rm
A}^2)^{1/2}}\over\pi G \sigma}>1\ .$$
On the other hand, in the presence of magnetic fields, azimuthal perturbations
are always unstable (albeit slowly if $\Bo$ is weak).
The unstable wavenumbers, obtained by
solving for $\lambda_{\rm A}^2$ in eq. (5b) and requiring
$\lambda_{\rm A}^2>0$, are such that
$k<J\sqrt{\gamma/(\gameff T_{\rm eq})}$ (the standard Jeans wavenumber), which
clearly can be smaller than the smallest available wavenumber in the
simulation. Thus, for practical purposes, our simulations with magnetic field
will be unstable in the azimuthal direction if the unstable scale is smaller
than the size of the integration box, i.e., $J^2 > (\gameff T_{\rm
eq}/\gamma)^{1/2}$. A similar condition for radial perturbations leads to
$J^2 > (\kappa^2 + \Bo^2 + \gameff T_{\rm eq}/\gamma)^{1/2}$.
If this latter condition is satisfied, the medium is then unstable
relative to both directions. When there is no magnetic field, the two
dispersion relations for the radial and azimuthal directions become identical,
and the instability criterion reduces to $J^2 > (\kappa^2 + \gameff T_{\rm
eq}/\gamma)^{1/2}$.

In the presence of shear, the above analysis is valid only for $t \sim 0$.
Confirming the results in E94, our numerical simulations indicate that,
for low shear rates, the magnetic field opposes radial
perturbations which tend to compress the field lines, while it
helps azimuthal perturbations since magnetic tension opposes
Coriolis spin-up.
For high shear, the magnetic field again stabilizes
azimuthal perturbations because they are sheared into the radial
direction before they have time to collapse (E94).
Finally, in the non-magnetic case, there is no distinction between
the radial and azimuthal directions, as stated before.

Table 3 shows the results of various simulations aiming at investigating the
combined effect of shear, magnetic fields, rotation and $\gameff$.
All runs have $J=.5$. Shown in
the Table are the relevant parameters of two sets of runs, without and with
magnetic fields. The value of $\kappa$ indicated in the Table is the minimum
over the integration box, as given by the shear profile.
The seventh column gives $C^2$,
an indicator of the linear instability of the runs, which we define as
the square of the ratio of the Jeans number to the appropriate
rotational or thermal pressure terms as discussed above. The eighth
column gives the actual behavior of the simulation, denoting by ``S'' and
``U'' stable and unstable runs, respectively. A rough estimate of the
actual growth rate of perturbations is given by the time --- in units of
$10^8$ yr --- taken by the
simulation to reach a peak density of 5 in code units. Note that in
Table 3 all runs have $\tilde Q_{\rm R} = \kappa/J > 1$, and are thus stable
to radial perturbations. Finally, the column labeled ``collapse'' gives a
description of the type and/or location of the clouds that form.

Note that in the present work, the initial perturbations are Gaussian with
random phases, and therefore contain both radial and azimuthal
components.
A first observation is that many of the runs require a long time before
collapsing, consistent with the fact that
the growth of the perturbations is not exponential but
oscillatory, due to the time-dependent nature of the linearized
equations (Elmegreen 1991a).
Indeed, a succession of epochs of growth and decay
is observed in our simulations before gravitational collapse
finally occurs in the unstable runs.

An interesting feature of the functional form we have chosen for
the shear (cf.\ eq.\ (2)) is that it allows representation of the
effects of various amounts of shear in a single simulation. The
profile crests have zero shear, while the nodes have maximum
shear. Recall that, as mentioned in \S 2.2, the shear has the same
sign as the actual Galactic shear ($d\Omega/dR < 0$) in only the
upper and lower quarters of the integration box.

We observe that all non-magnetic runs in Table 3, except run 45, should be
stable according to the linear criterion based on $C^2$. However, the
linear behavior is seen to be modified by the presence of shear,
as exemplified by runs 39 and 105, which differ only by
the presence of shear in the former. Although both should be stable, since $C^2
< 1$, run 39 actually forms clouds in a relatively short time. All of
these runs exhibit cloud formation in the regions where $d\Omega/dR < 0$ (which
minimizes $\kappa$), a reflexion of the original linear criterion.
Note that the run which exhibits cloud formation on the shortest
timescales (run 45) is the only one that
is unstable according to the linear criterion. Moreover, this run exhibits
clouds formation without preference for specific regions in the integration box
because it has $\kappa =0$.

In the presence of a magnetic field, all runs in Table 3 should be unstable
according to the linear criterion, and indeed are. However, the stabilizing
effect of the magnetic field in the presence of high shear (E94) can be seen in
the fact that the simulations with shear tend to form clouds preferentially in
the regions of zero shear. Even though the linear instability criterion does
not
include the value of the magnetic field, nevertheless it can be observed that
runs with larger $\Bo$ tend to form clouds more rapidly (compare runs 57
and 47; and runs 56 and 43). This result exemplifies the inhibition
of Coriolis spin-up by the
magnetic field (magnetic braking). Also, comparing runs 42 and 43, which are
identical except for their value of $\gameff$, we see that the run with zero
pressure gradient collapses faster, again a reflection of the original linear
criterion. Finally, note that the runs labeled ``azimuthal'' in the
``collapse'' column have no shear, leading to a collapse simply along field
lines, and forming a radially-oriented cloud.

These results are exemplified in figs.\ 1a and 1b, in which typical
unstable non-magnetic (fig.\ 1a) and magnetic (fig.\ 1b) runs are respectively
shown at late times in their evolution. In fig.\ 1a, a large
elongated cloud is seen to form in the upper and lower quarters of
the domain (recall the boundary conditions are periodic), where
$d\Omega/dR < 0$. In fig.\ 1b, clouds are seen to form at the
crests of the shear profile, where $d\Omega/dR = 0$.

Finally, non-magnetic runs with shear (e.g.\ run 41) initially
exhibit an expulsion of material from the central regions of the
box, where $d\Omega/dR > 0$, indicating that the classical
Rayleigh instability is at work there.

\bigskip
\bigskip

\centerline{4. MAGNETIC EFFECTS IN THE FULLY TURBULENT REGIME}
\medskip
\centerline{4.1 {\it Influence of star formation on
magnetic field dynamics}}

This section is devoted to a study of the interplay between
magnetic field and star formation (SF) in the fully turbulent
regime.  We will start with the influence that the stellar forcing
has on the magnetic field dynamics.

Since our simulations are two-dimensional, we cannot expect any
dynamo mechanism. Growth and maintenance of magnetic fluctuations
are nevertheless observed in our system. In order to understand
the mechanism at play, it is convenient to write the equation for
the evolution of the quantity ${\cal P}\equiv\int\rho A^2 d^2 \x$
where $A$ is the vector potential defined by the relation
$\B=\nabla \times (A \e_z)$.  After splitting the magnetic field
$\B$ into its constant component $\vBo=\Bo \e_x$ and its
fluctuating part $\b=\nabla \times (a \e_z)$, we can write an
equation for the fluctuating potential
$$ {\partial a \over
\partial t}=(\u \times \B) \cdot \e_z+\eta (-1)^{n+1}\nabla^{2n}
a=-\u\cdot \nabla a +(\u \times \vBo)\cdot\e_z+\eta
(-1)^{n+1}\nabla^{2n} a, $$
where $n=1$ for standard MHD
and $n=4$ in our simulations.  Assuming the fields are
periodic in the domain or vanish at the boundaries, one obtains
for the time evolution of ${\cal P}$:
$$ {\partial\over \partial
t}\int \rho a^2 d^2 \x=2\int\rho a (\u \times \vBo)\cdot \e_z d^2
\x-2\eta\int\nabla^n (\rho a)\cdot (\nabla^n a) d^2 \x $$
$$ =-2
\Bo\int\rho a u_y d^2 \x -2\eta\int\nabla^n (\rho a)\cdot (\nabla^n
a) d^2 \x.  \eqno (6) $$
As is well known (Moffatt, 1978) when $\Bo=0$ and $\rho$
is constant, the quantity ${\cal P}$ decreases to zero, and thus
magnetic fluctuations die away, whatever forcing is applied on the
velocity field.  When density fluctuations are small enough, the
contribution of dissipation to the evolution of ${\cal P}$
is again certainly negative. In the general case, one
cannot assert if ${\cal P}$ will decrease monotonically,
%
%
although one expects that in general the effect of dissipation will be a
cumulative decrease in magnetic energy.
%
%
Let us
now consider the case $\Bo\neq 0$. When the system is forced by
Alfv\'en or magnetosonic waves, the first term in the right-hand side of
eq.\ (6) still vanishes and thus magnetic field fluctuations
cannot be maintained.  This can be seen if one takes for $a$,
$u_y$ and $\delta \rho$ small disturbances of the equilibrium
state $a=0$, $u=0$, $\rho=1$, proportional to the eigenvectors of
the linearized equations. When these disturbances have an
oscillatory behavior (stable case), $\delta\rho$ and $u$ have the
same parity, opposite to that of $a$, so that $\int (1+\delta
\rho)a u\ d^2 \x=0$.  When these disturbances grow or decay on the
other hand (unstable case), $a$ and $u$ have the same parity,
opposite to that of $\delta \rho$, so that the first part of the
integral, $\int a u\ d^2 \x$, is nonzero. The first term on the
r.h.s. of eq.\ (6) is positive and can thus balance or even
overcome dissipation.  It is precisely the situation when a
single cloud contracts or relaxes.  The fields $\rho$, $a$, $u_y$
are schematically depicted in fig.\ 2a for a magnetosonic wave and
in fig. 2b for a cloud contraction.  This mechanism is probably as
efficient as a dynamo to sustain magnetic fluctuations, but a
dynamo is still needed to generate the locally constant field
$\Bo$ (the constant mode $\Bo$ is dynamically disconnected from the other
modes).

In order to test numerically the mechanism for the generation of
magnetic fluctuations, we have performed the series of runs
denoted $12$, $13$, $14$ in Table 2 for which the initial
fluctuations of the magnetic field are set to $\delta B=0.01 $ and
the constant field $\Bo$ takes respectively the values $0$, $0.03$
and $0.3$. Figure 3 displays the fluctuating magnetic energy as a
function of time for the three runs. It is very clear that we
indeed have a growth of $\int (\delta B)^2 d^2 \x$ when $\Bo\neq 0
$. The values of $\int (\delta B)^2 d^2 \x$ reached during the run
are also increasing with $\Bo$. The fluctuating magnetic energy
and the density fluctuations are correlated when $\Bo$ is nonzero,
whereas they are not when $\Bo=0$ (not shown).

A few additional points are worth noting concerning the constant field $\Bo$.
First, its presence is a
regularizing factor in inviscid incompressible MHD (Bardos, Sulem \& Sulem,
1988).
Second, it
also removes neutral points (probably not all of them if it is too
weak) and thus leads to a decrease of magnetic dissipation. Finally, it has
also been shown that the presence
of a constant field $\Bo$ hinders small scale turbulence
(Shebalin, Matthaeus \& Montgomery, 1981). In compressible MHD turbulence
permeated by a uniform magnetic field, the
solenoidal enstrophy is much smaller than when $\Bo=0$, whereas
its compressible counterpart or density fluctuations remain
unaffected (A. Broc, 1993).

In every MHD simulation, we observe a tendency for the typical
scale of the magnetic fluctuations to grow in time. This effect
can be observed in figs.\ 4a and 4b, which respectively display the
magnetic spectra at times $t=.65 \times 10^8$ and $t=2.6 \times
10^8$ yr for run 15. Note the growth of the $k=1$ mode for the
$b_y$ component.  A study in three dimensions would be required to
test whether this mechanism persists or is a consequence of the
two-dimensional assumption.  But let us mention that in the
three-dimensional incompressible case, there is an inverse cascade
of magnetic helicity, with the magnetic energy following to a
lesser extent (Pouquet, Frisch \& L\'eorat, 1976; Meneguzzi,
Frisch \& Pouquet, 1981; Horiuchi \& Sato, 1986, 1988).

Figure 5 also displays spectra of the total magnetic energy
together with those of the solenoidal and compressive components
of the velocity field.  Note that the magnetic spectrum is almost
in equipartition with the solenoidal spectrum, as already observed
in two-dimensional
compressible MHD simulations (Pouquet, Passot \& L\'eorat, 1991;
Dahlburg \& Picone, 1989; Picone and Dahlburg, 1991).  Contrary to the
case of a decaying MHD compressible turbulence, we do not observe
a systematic growth of $\u \cdot \B$ correlations.  On the other
hand it is quite clear that for finite but moderate values of
$\Bo$, this correlation oscillates noticeably, the extrema
increasing with $\Bo$. When $\Bo$ is zero and the wavenumber of
magnetic fluctuations $k_B$ is initially equal to $4$, the correlation
actually
goes steadily to zero.  The temporary growth of $\u \cdot \B$
correlations is probably due to a transfer of magnetic energy
created by clump contraction, to large scale Alfv\'en waves along
$\Bo$. These results are markedly different when $\Bo$ is very
large because of a strong confinement of the clouds (see below).
A phenomenon which is also common to our present and former MHD
simulations (Pouquet et al.\ 1991),
is that {\it the global compressibility of the medium (as measured by
the ratio of compressible to total kinetic energy) increases with
the intensity of the fluctuating magnetic energy}.

It is interesting to note that the topology of the magnetic field,
mainly parallel to the local shear, is actually enforced by the
shear itself. A run with the uniform component of the magnetic
field initially perpendicular to the direction of the shear,
evolves so that the magnetic field becomes aligned with the shear
profile.

{\it The field topology is almost uniform in intercloud regions and
very turbulent in cloud complexes undergoing star formation} (fig.\ 6). The
magnetic field has actually an important dynamical effect in these
regions, favoring clumpiness and fragmentation.  On the other
hand, at its fiducial value, it seems rather inefficient in
intercloud regions to influence cloud motions. We indeed observe
clouds propagating almost in any direction with respect to the
magnetic field.  The turbulent character of the magnetic field
inside clouds is also possibly at the origin of the lack of
correlation between $B^2$ and $\rho$ in these regions, as
indicated by fig. 7b which shows a scatter plot of the square of the
magnetic field strength as a function of density for the complex encircled
in fig. 7a.

Finally, the typical values of the field strength within clouds is
$\sim 12 \mu$G, but can possibly reach $30\mu$G. This can be compared
to the value in the ICM of $\sim3 \mu$G. As opposed to quasi-stationary models,
which cannot reach such high contrasts between cloud and intercloud field
strengths (Mc Kee et al. 1993), the present turbulent model is capable of
producing realistic values of the magnetic field strength within clouds.

\medskip \centerline{4.2 {\it Influence of the magnetic field on
cloud and star formation}}

The second part of this section is devoted to the influence of
magnetic fields on the formation of density clumps (eventually
leading to star formation).  As in the linear regime, a constant
field $\Bo$ hinders cloud condensations and thus star formation
when it is small (super-Alfv\'enic motions) but favors it when it
is large (sub-Alfv\'enic motions).  In the case of even much
larger values of $\Bo$, star formation gets inhibited again.
These results are summarized in fig. 8 which displays the total
star formation (SFR integrated over the total lapse time of the
simulations) for runs with different values of the constant
magnetic field $\Bo$ but otherwise identical (runs 53, 6, 19, 54, and
61 of Table 2). If the uniform field $\Bo$ is relatively weak, clumps get
significantly disrupted by the shear -- originating either from
differential rotation or from turbulence -- before they can
contract along the magnetic field; further contraction, that has
to occur across the magnetic field, is then hindered by magnetic
pressure.  On the other hand if the uniform field is relatively strong, it
can act against the stabilizing action of the Coriolis force
through the magnetic braking mechanism, and clump contraction can
then occur along the field before shear can disrupt the
condensation.  This trend is valid even when the magnetic field
fluctuations are at smaller scale (see runs 11, 15 and 5 in Table 2).

For even larger values of $\Bo$,
another effect comes into play that we call the ``pressure
cooker'' effect. Shells formed by stellar heating cannot expand as
much when the magnetic field is strong, resulting in confining
clouds in complexes or very thick filaments. Inside these
complexes, star formation induces a strongly turbulent state where
smaller roundish clouds get formed (see fig. 9 for run 61).
Figure 10 also shows this effect by comparing three runs with different
values  of the initial uniform magnetic  field component (runs 53, 19 and 61
in Table 2).
However, with such an
inhibition of shell expansion, stellar activity
is {\it globally} reduced (self-propagating SF is not as
effective); also reduced are both the compressibility and the
amount of $\u \cdot \B$ correlations since Alfv\'en waves cannot
get efficiently excited.
%
%
Now, the cloud complexes have a tendency
to follow the shear. This ``pressure cooker'' effect is present
even at smaller values of $\Bo$ although it is less obvious.
Expansion motions tend to amplify magnetic fluctuations
perpendicular to the velocity, and as a result these motions get
decelerated due to the influence of the tension of the amplified
field lines. In these regions, the field lines are perpendicular to the
density gradients, i.e.\ parallel to the density features (fig. 6).

In general, one can say that
the effects of a large uniform field are somewhat opposite to those of a
large fluctuating field.

Planar shocks propagating perpendicular to a constant magnetic
field have a reduced compressibility, compared to the non-magnetic
case, and if the pre-shock state is at rest, the shock must
propagate at $U_s>(c^2+v_A^2)^{1/2}$.  If the field gets
amplified beyond a certain value due to nonstationary expanding
motions, the shock will not be able to propagate and will give
rise to a magnetosonic wave. The shell will then stop to propagate
as an ``entity'', and to collect matter on its path, and the motion of
the resulting cloud will become a wave propagation.
In the simulations we observe that shells propagate at speeds of the order
of 8 km s$^{-1}$ while the typical sound speed in the ICM is $\sim 12$
km s$^{-1}$ and
the Alfv\'en speed $\sim 30$ km s$^{-1}$.
Other effects take place when the field is curved. As mentioned above,
the field line tension tends to oppose expanding motions, while the Lorentz
force, which is radial and oriented inward, promotes converging
motions and generates blobs.  As a result, planar structures are
less likely to subsist in the magnetic case and we actually
observe more roundish structures when the magnetic field is
stronger and/or at smaller scales.  Also, an expanding shell will
break up more easily in presence of a magnetic field due to
the Rayleigh-Taylor (and possibly the Parker) instability.
%
%

Another effect of the magnetic field is linked to the scale of its
fluctuations.  Star formation is reduced when the field
fluctuations are on smaller scale (compare runs 53, 19 and 54 with runs
11, 15 and 5 in Table 2). This effect is the result of
the magnetic pressure acting at a comparable scale as the one of
the fluid motions leading to clump contraction. In a run initiated
with $\kb=4$ (as opposed to the standard case with $\kb=1$), the
inverse cascade mechanism gradually enhances the large-scale
magnetic field, and consequently star formation increases with
time in this run.

Star formation is also proportional to the intensity of the
magnetic fluctuations $\delta B$, the more so when these are on
large scale, because in that case, they locally act as a constant
magnetic field (compare runs 8 and 7 with runs 19 and 5 in Table 2).

Before closing this section, in summary we emphasize that
in the turbulent runs, even though we partly recover the general trends
of the linear stability analysis, it is however clear that the
formation, dynamics and morphology of clouds are
dominated by turbulent energy injected from stars, rather than by the combined
linear instability.

\bigskip
\bigskip
\centerline{5. GENERAL BEHAVIOR OF A FULLY TURBULENT SIMULATION}
In this section we briefly describe some noteworthy features of run 28, a
simulation with fiducial values of the parameters, using $512 \times
512$ grid points (see Table 2). According to our choice of parameters,
one pixel corresponds to a size of roughly 2 pc. Although our viscosity is
kept at the minimum compatible with the stability of our numerical scheme,
it damps velocity features at scales less than about 10 pc.

Figures 11a, b and c respectively show the density, pressure and temperature
for this run at time $.72 \times 10^8$ yr. Other views of
this run at different times are given in figs.\ 6 and 7a. In fig. 11a, we
identify
structures that can be called ``giant complexes'', with sizes of several
hundreds of pc, ``complexes'', with sizes $\sim 100$ pc, and ``clouds'', with
sizes af a few tens of pc. However, we stress that this classification is
rather arbitrary, since in reality there is a continuum of density structures
that can be identified by thresholding the density at continuously varying
levels. This fact led us in Paper I to use a continuous filling factor
function.  For the MHD simulations, this point and
statistical properties of the clouds will be discussed in a future paper.

One of the most striking features of the density structures in the simulation
is their amorphous character, none of the clouds being nearly circular, nor
even elliptical. Moreover, density structures are hierarchically nested, as
observed in real interstellar clouds (see, e.g., Scalo 1985). This phenomenon
was also observed in non-magnetic simulations (\VS, 1994, Paper I).

Extremely long and thin filaments of density and pressure (with low contrasts)
are observed. The filaments are correlated with
jumps of the velocity and magnetic field (compare
figs.\ 11a and b with the magnetic field components shown in
figs.\ 12a and 12b).
This is also apparent in both the vorticity and the current (not shown).
Similar filamentary
structures obtain in the non-MHD runs of Paper I, as well as in the simpler
non-rotating decay runs of compressible turbulence
(Passot et al., 1988).

Star-forming (``HII'') regions have sizes of a few tens of pc, thus
corresponding to the largest observed sizes for HII regions in real galaxies.
In our simulations, these regions cannot be much smaller than this because of
the Gaussian smoothing used for numerical reasons (Paper I, \S 2.2).
``HII'' regions in the simulations are most easily observed in the pressure and
temperature images, as seen in figs.\ 11b and 11c. A particularly active
star-forming region is seen in fig.\ 11b near the upper-right corner, and in
fact this region is somewhat evocative of images of the neighborhood of
30 Dor in the Large Magellanic Cloud.

Global gravitational contraction of a giant complex appears to occur in the
upper right quadrant, as can be seen in the density field for run 28 at times
$t=.42 \times 10^8$yr (fig.\ 6), $t=.72 \times 10^8$yr (fig.\ 11a), and $t=.91
\times 10^8$yr (fig.\ 7a). However, generalized collapse is halted by
SF activity, which
increases the turbulence in the highest-density regions of the complex.
Also, the kinetic and magnetic energies in various clouds appear to be within
a factor of 5 from each other, suggesting rough equipartition.
A detailed study of the energetics of the clouds and
complexes in run 28 will be presented in a future paper.

\bigskip
\bigskip
\centerline{6. CONCLUSIONS}

\centerline{6.1. {\it Summary}}

In this paper we have presented an extension of the model
introduced in Paper I, now incorporating magnetic fields, Galactic
disk rotation and a density-dependent diffuse heating which
results in a softer equation of state.  The behavior of the model
was studied both in the linear and nonlinear regimes.  In the
linear regime, the model is well described by the linear theory
developed by Elmegreen (1991a, E94; see also Elmegreen 1991b).
At low shear, a weak field stabilizes
the medium by opposing collapse of radial perturbations, while a strong field
is destabilizing by preventing Coriolis spin-up of azimuthal perturbations
(magnetic braking). At high shear, azimuthal perturbations are sheared into the
radial direction before they have time to collapse, and the magnetic field
becomes stabilizing again.
In the absence of magnetic field,
the problem reduces to the classical Toomre (1964) criterion, slightly modified
for our 2D model.

In the nonlinear regime, a variety of interesting effects are
present, many of them unforeseeable through linear analyses: \par

\item{}
An amplification mechanism for the fluctuating magnetic field has been
identified, that allows the maintenance of magnetic energy over the
long time evolution of the interstellar medium. It requires the presence of a
constant component $\Bo$ and is effective for contracting or expanding motions,
such as the expanding shells generated by stellar heating.\par

\item{}
The fluctuating magnetic field is generated at the scale of
a cloud and is then transferred  to larger scales.\par

\item{}
The magnetic field dynamics clearly does not reduce to the propagation
of waves. The field participates to the global magneto-hydrodynamic turbulence,
as demonstrated by the magnetic spectra, which exhibit
a developed inertial range.\par

\item{}
The importance of the magnetic field on the global dynamics is complex.
It depends both on its topology and its intensity compared to the
shearing motions. A strong magnetic field confines matter in big complexes
where smaller roundish clumps are formed, globally inhibiting star
formation. The field is turbulent inside the clouds and straight in
intercloud regions.\par

\item{}
No unique correlation is found between the density and
the magnetic field intensity
probably due to its turbulent character. However, its orientation
at the edge of the clouds tends to be perpendicular to the density
gradient. Also,
the magnetic field strength within clouds is roughly a factor of 4
larger than in the ICM with excursions of up to factors of 10.
This suggests that the field in clouds is amplified through flux freezing by
the
same collisions of gas streams that form the clouds
(Hunter et al.\ 1986; Elmegreen 1993; Paper I).\par

\item{}
The velocity of the expanding shells rapidly becomes sub-Alfv\'enic. \par

\bigskip
\centerline{6.2. {\it Discussion of limitations}}

Some of the most important
limitations of this work are the relatively low effective resolution,
the two-dimensionality of the simulations, and the omission of supernovae.
In this section we briefly discuss their possible consequences.

As mentioned in \S 5, viscosity damps velocity features at scales smaller than
3--5 pixels ($\sim$ 6--10 pc at a resolution of 512 grid points per
dimension). Additionally, the mass diffusion smooths out density fluctuations
at comparable scales. These scales correspond to those of sizable molecular
clouds.
Thus, the simulations are incable of resolving the structure within the
clouds of those sizes, the applicability of the model being restricted
to structures ranging from superclouds to large individual molecular clouds.
However, the
usage of a hyperviscosity scheme with a high power of the Laplacian guarantees
that features in this range of scales are
virtually unaffected by viscosity, and are thus fully turbulent.
Moreover, note that in this paper we have focused on problems that do not
require a very high resolution to deal with, as evidenced by the fact that only
one run of those reported in Table 2 has a resolution of 512 grid points per
dimension, while all others have 128.

The two-dimensionality of the simulations has several effects.
Among them are the absence of vortex stretching,
and the modification of
the gravitational potential leading to a force proportional to the inverse
of the distance between clouds. Other effects are also mentioned in \VS\ (1994)
and Paper I.
Moreover, a dynamo effect is obviously absent from our
simulations.  However, we have identified an alternative mechanism
by which expanding bubbles can amplify the magnetic energy through
amplification of the magnetic potential, provided a uniform
component of the field is present.

Another important consideration is that
the two-dimensionality of the simulations implies
the existence of an inverse (from small to large scales) energy cascade which
is not present in three dimensions
(Kraichnan 1967), and therefore a reduced rate of energy
dissipation at small scales compared to the 3D case (see, e.g., Lesieur 1990).
Note, however, that this
does not lead, for example, to excessively turbulent regimes.
In purely-hydrodynamic incompressible turbulence with random extended forcing,
the inversely-cascading energy organizes itself into
large-scale vortex pairs which exist within a normally turbulent background
(McWilliams 1984).
Furthermore, in the case of our compressible simulations with a pointwise
compressible forcing, the latter (stellar heating) produces expanding shells
that create turbulence at large scales, although without the formation of
large-scale vortices. This process is different from an
inverse cascade and must be present in 3D as well as in 2D, decreasing the
difference between the two cases.

Supernovae (SNe) are omitted from the present work in part because
new code development would be required. It is not clear whether their
presence will
have (or not) a sizable effect in particular on the density and
temperature of the ICM. Such an omission results in the absence in
our models of a hot ($T \sim 3 \times 10^5$--$10^6$ K, $\rho \sim
10^{-2}$ cm$^{-3}$) gas phase.  Under the classical picture of
McKee \& Ostriker (1977), this hot gas occupies the vast majority
of the volume (filling factor $\sim 1$), and in this framework our
warm ICM ($T \sim 10^4$ K, $\rho \sim 10^{-1}$ cm$^{-3}$) ought to
be be essentially replaced by the hot gas. On the other hand,
recent numerical results of Slavin \& Cox (1992, 1993) suggest
that the filling factor of the hot gas may be as low as 20\%. In
this case, our results would only be modified by the inclusion of
mostly isolated bubbles of hot gas with a relatively small filling
factor. Which view is more realistic is currently a matter of
debate, and depends on a variety of issues, such as whether
a supernova remnant
breaks out of its high-density surroundings before it has had time
to cool off (e.g., Mac Low \& McCray 1988; also G. Garc\'\i
a-Segura, private communication), and whether the magnetic field
forces the surrounding shell to rebound (Slavin \& Cox 1992). Our
finding that a ``pressure cooker'' effect reduces shell expansion
for our model HII regions seems to support the Slavin \& Cox
picture, but the degree to which the low-density gas in our
simulations will change upon introduction of SNe cannot be
assessed precisely.
The simulations of Rosen \& Bregman (1994) do include SNe and
contain a sizable fraction of hot
gas, but since their SF scheme is smooth in space and time, significant
differences may arise upon consideration of a more realistic,
spatially-discrete, thresholded
SF scheme like the one used in the present paper.
Moreover, their calculations do not include the magnetic field.

In any case, in spite of our neglect of SNe, it is
likely that the morphology and energetics {\it of the clouds} will not
be dramatically affected even if the hot phase is pervasive, as
its pressure should be comparable to that of our warm ICM and the
dynamics induced by expanding shells is already present in our
simulations and with comparable total kinetic energies (note that
the total kinetic energy inputs to the ISM from OB ionizing
radiation, winds, and from SNe are of comparable magnitude (Abbott
1982; Garc\'\i a-Segura, Mac Low and Langer 1995)).  On the other
hand, the inclusion of forcing from
SNe and OB winds may result in the medium being
even more turbulent, and the mechanism of cloud formation by
turbulent density fluctuations exemplified in Paper I might become
even more important.
Work considering these effects is in progress (Gazol et al.\
1995), in which we will address questions such as the filling factors of the
various gas phases in the midplane of the Galactic disk and the longevity of
the $10^6$ K gas.

\bigskip
\bigskip
We gratefully acknowledge comments from and/or
discussions with Steve Balbus, John Dickey, Bruce Elmegreen,
Edith Falgarone, George Field, Guillermo Garc\'\i a-Segura, Carl
Heiles, Jean-Loup Puget, Alex Rosen and Ellen Zweibel.
The numerical calculations
were performed on the Cray C98 of IDRIS (CNRS), France, and the Cray Y-MP
4/64 of DGSCA, UNAM, M\'exico. This work has received partial
financial support from EEC Human Capital Network Grant ERBCHRXCT930410 to T.\
P. and A.\ P., and grants DGAPA IN101493, CRAY/UNAM SC000392,
as well as
a visiting-astronomer position at the Observatory of Nice to
E.V.-S.

\bigskip
\bigskip
\parindent = 0 pt
\centerline{REFERENCES}

Abbott, D. C. 1982, ApJ, 263, 723

Babiano, A., Basdevant, C., Legras, B., \& Sadourny, R. 1987,
J. Fluid Mech., 183, 379

Binney, J., \& Tremaine, S. 1987, Galactic Dynamics (Princeton: Univ. Press)

Canuto, C., Hussaini, M. Y., Quarteroni, A., \& Zang, T. A. 1988,
Spectral Methods in Fluid Dynamics (Berlin: Springer-Verlag)

Chandrasekhar, S. 1961, Hydrodynamic and Hydromagnetic Stability (London:
Oxford University Press), p. 589

Crutcher, R. M., Kazes, I., \& Troland, T. H. 1987, 1987, A\&A, 181, 119

Dalgarno, A., \& McCray, R. A. 1972, ARAA, 10, 375

Elmegreen, B. G. 1991a, ApJ, 378, 139

Elmegreen, B. G. 1991b, in The Physics of Star Formation and Early Stellar
Evolution, ed. C. J. Lada \& N. D. Kylafis (Dordrecht: Kluwer), p. 35

Elmegreen, B. G. 1993, ApJ, 419, L29

Elmegreen, B. G. 1994, ApJ, 433, 39

Falgarone, E., \& Puget, J. L. 1986, A\&A, 162, 235

Falgarone, E., Puget, J.-L. \& P\'erault, M. 1992, A\&A, 257, 715


Franco, J., \& Cox, D. 1986, PASP, 98, 1076

Gammie, C. 1994, ApJ (submitted)

Garc\'\i a-Barreto, J. A., Burke, B. F., Reid, M. J., Moran, J. M., Haschick,
A. D., \& Schilizzi, R. T. 1988, ApJ, 326, 954

Goldreich, P. \& Lynden-Bell, D. 1965, MNRAS, 130, 7

Goodman, A. A. 1991, in Atoms, Ions and Molecules: New Results in Spectral Line
Astrophysics, ed. A. D. Haschick \& P. T. P. Ho (San Francisco: ASP), 333

Heiles, C. 1990, ApJ, 354, 483

Heiles, C., Goodman, A. A., McKee, C. F., \& Zweibel, E. G. 1993, in Protostars
and Planets III, ed. E. H. Levy \& J. I. Lunine (Tucson: Univ. of Arizona
Press), 279

Kutner, M. L. 1994, BAAS, 26, 1397

Mac Low, M.-M. \& McCray, R. 1988, ApJ, 324, 776

McKee, C. F., \& Ostriker, J. P. 1977, ApJ, 218, 148

McKee, C. F., Zweibel, E. G., Goodman, A. A., \& Heiles, C., in Protostars and
Planets III, ed. E. H. Levy \& J. I. Lunine (Tucson: Univ. of Arizona
Press), 327

McWilliams, J. 1984, J. Fluid Mech., 146, 21

Meneguzzi, M., Frish, U., \& Pouquet, A. 1981, Phys. Rev. Lett. 47, 1060

Mouschovias, T. 1976a, ApJ, 206, 753

Mouschovias, T. 1976b, ApJ, 207, 141

Myers, P. C., \& Goodman, A. A. 1988a, ApJ, 326, L27

Myers, P. C., \& Goodman, A. A. 1988b, ApJ, 329, 392

Pouquet, A., Passot, T., \& L\'eorat, J. 1991, in Fragmentation of
Molecular Clouds and Star Formation, IAU Symp. 147, ed. E. Falgarone,
F. Boulanger \& G. Duvert (Dordrecht: Kluwer), 101

Pudritz, R. E., \& G\'omez de Castro, A. I. 1991, in Fragmentation of Molecular
Clouds and Star Formation, ed. E. Falgarone, F. Boulanger, \& G. Duvert
(Dordrecht:Kluwer), 317

Rand, R. J., \& Kulkarni, S. R. 1989, ApJ 343, 760

Raymond, J. C., Cox, D. P., \& Smith, B. W. 1976, ApJ 204, 290

Rosen, A., Bregman, J. N., \& Norman, M. L. 1993, ApJ, 413, 137

Rosen, A., \& Bregman, J. N. 1994, BAAS, 25, 1394

Shore, S. N. 1989, in Encyclopedia of Astronomy and Astrophysics, ed. R. A.
Meyers (San Diego: Academic Press), 183

Shu, F. 1992, Gas Dynamics (Mill Valley: University Science Books)

Slavin, J. \& Cox, D. 1992, ApJ, 417, 187

Slavin, J. \& Cox, D. 1993, ApJ, 417, 187

Struck-Marcell, C. \& Scalo, J. 1984, ApJ, 277, 132

Toh, S., Ohkitani, K., \& Yamada, M. 1991, Physica D, 51, 569

Toomre, A. 1964, ApJ, 139, 1217

Trimble, V. 1990, in Galactic and Intergalactic Magnetic Fields, ed. R. Beck,
P. P. Kronberg, \& R. Wielebinski (Dordrecht: Kluwer), 29

Troland, T. H. 1990, in Galactic and Intergalactic Magnetic Fields, ed. R.
Beck, P. P. Kronberg, \& R. Wielebinski (Dordrecht: Kluwer), 293

V\'azquez-Semadeni, E., Passot, T. \& Pouquet, A. 1995, ApJ

Wielebinski, R., \& Krause, F. 1993, A\&A Rev., 4, 449

Zweibel, E. 1987, in Interstellar Processes, ed. D. J. Hollenbach \&
H. A. Thronson (Dordrecht: Reidel), 195

\bigskip
\bigskip

\centerline{FIGURE CAPTIONS}

\bigskip
Fig. 1. a) Contour plot of the density field of run 39, a
non-magnetic run in the linear regime, at $t=2.47 \times 10^8$ yr.
$\Omega$ points upwards perpendicular to the plane of the page.
A large cloud has formed in the
region where $d\Omega/dy<0$ (the upper and lower quarters of the domain; recall
the boundary conditions are periodic). Maximum density for this plot is
$\rho_{\rm max} \sim 5$. After this time, this cloud still goes through several
oscillations before collapsing. b) Contour plot
of the density field of run 43 at
$t=2.61 \times 10^8$ yr. This is a magnetic run, also in the linear regime. In
this case, clouds form at the crests of the shear profile given by eq. (2),
where $d\Omega/dy=0$. Maximum density for this plot is $\rho_{\rm max} \sim
3$.

\medskip
Fig. 2. Sketches of the density $\rho$ ({\it dash-dotted line}),
velocity $u$ ({\it long-dashed line}) and
magnetic potential $a$ ({\it solid line}) for a
magneto-sonic wave ({\it top}) and for a
cloud compression ({\it bottom}) along the $y$-axis, perpendicular to $\Bo$.
For waves, $\rho$
and $u$ have the same parity, opposite to that of $a$, whereas for the
compression $a$ and $u$ have the same parity, opposite to that of $\rho$.

\medskip
Fig. 3. Evolution of the fluctuating component $({\bf B} - {\bf B}_\o)^2$ of
the magnetic field for three runs with different initial values of the uniform
component of the magnetic field $\Bo$. {\it Solid line}: run 12, $\Bo=0$. {\it
Dotted line}: run 13, $\Bo=.15\ \mu$G. {\it Dashed line}: run 14,
$\Bo=1.5\ \mu$G. Both the amplitude of the fluctuations and their time
derivative are seen to increase with $\Bo$.

\medskip
Fig. 4. Magnetic spectra of run 15 (initial
magnetic centroid $k_B=4$) at two
different times: $t_8=0.65\times 10^8$ yr ({\it top}) and
$t_8=2.6\times 10^8$ yr ({\it bottom}).
Solid lines correspond to $B_x$ and dotted lines to $B_y$.
Note that the energy in both components of the magnetic field
is transferred to lower modes as time increases
(particularly so for the $B_y$ component).

\medskip
Fig. 5. Velocity and magnetic spectra for run 28 at $t=2.6\times 10^8$ yr. {\it
Solid line}: total magnetic spectrum. {\it Dotted line}: solenoidal velocity
spectrum. {\it Dashed line}: compressible velocity spectrum. Note that the
magnetic and the solenoidal spectra are within factors of a few from each other
at all scales.

\medskip
Fig. 6. Density and magnetic fields for run 28 at $t=0.42 \times 10^8$. The
density grey scale is logarithmic, and saturates at $\rho=40$ cm$^{-3}$. The
arrow at the bottom right indicates a field strength of 30 $\mu$G. Note the
strong magnetic turbulence inside clouds and the rather smooth character
of the magnetic field in the ICM. Regions of alignment of the magnetic field
and density features can be seen in the filament in the upper right corner.
However there are also regions where the magnetic field is perpendicular
to the density features such as the lower portion of the same cloud
and also the cloud near the center of the lower left quadrant.

\medskip
Fig. 7. (a) Density field for run 28 at $t=9.1 \times 10^7$ yr
showing a circular complex of radius 30 pixels ($\sim 60$pc). The density grey
scale is as in fig.\ 6. (b) $B^2$ vs.\ $\rho$ for the
circular region of fig.\ 7a. Spherically symmetric compressions should give
a linear relationship. The scatter diagram indicates that there is
no preferred compression geometry.

\medskip
Fig. 8. Time integral of the star formation (SF) rate for various runs as
a function of the initial value of the uniform component of the
magnetic field $\Bo$. From smaller to larger values of $\Bo$, the points
represent runs 53, 6, 19, 54 and 61. At small
$\Bo$, SF is inhibited because $\Bo$ is small enough not to counteract magnetic
braking, but is able to prevent radial collapse of sheared
condensations. Intermediate values of $\Bo$ counteract magnetic braking and
thus
promote SF. Very large values of $\Bo$ inhibit SF again because the magnetic
field rigidifies the medium.

\medskip
Fig. 9. Contour plot of the density field for run 61 with magnetic field
vectors superimposed. The uniform magnetic field strength $\Bo$ for this run
is $10\mu$G. At this value of $\Bo$ (about six times the fiducial value), the
magnetic field is almost unperturbed by the fluid motions but the clouds
exhibit more roundish shapes. The magnetic field fluctuations, even though
they are small compared to $\Bo$, occur mainly inside the clouds.

\medskip
Fig. 10. Contour plots of the density field at $t=5.2 \times 10^6$ yr
for three runs with progressively larger values of $\Bo$, but otherwise
identical. From left to right, run 53 ($\Bo = 0$), run19 ($\Bo = 1.5 \mu$G),
and
run 61 ($\Bo=10 \mu$G). Note the tendency
towards more roundish structures as $\Bo$ increases.

\medskip
Fig. 11. Grey scale images of (a) density, (b) pressure and (c) temperature
fields of run 28 at $t=.72 \times 10^8$ yr.
The bright spots within the large cloud complex in the upper right are
star-forming regions, or ``HII regions''. The grey scale in a) and b) is
logarithmic. The density grey scale saturates at $\rho=40$ cm$^{-3}$ and the
pressure scale saturates at $P= 5$ in code units
($8.3 \times 10^4$ cm$^{-3}$ K).
Note the globally low pressure contrast, with typical intercloud
values ranging between 3,300 and 6,700 cm$^{-3}$ K, and typical cloud values
around 10,000 cm$^{-3}$ K. Exceptions are the
``HII'' regions, where the pressure reaches $1.8 \times 10^5$ cm$^{-3}$ K,
and which are strongly saturated in this image.
In c), the grey scale is linear and spans a temperature range of 0 to 20,000 K.

\medskip
Fig. 12. Grey scale images of the $x$- ({\it a}) and $y$- ({\it b}) components
of the magnetic field for run 28 at $t=.72 \times 10^8$ yr. The grey scale
ranges from $-7.5$ to 7.5 $\mu$G. The field is most
turbulent in complexes and clouds. Note the discontinuities in both components
of the field, which are coincident with the long, thin filaments observed in
the density and pressure fields (figs.\ 11a and b).

\end